\documentstyle[aps,prl,epsfig,multicol,floats]{revtex}

\begin{document}
\draft


\title{Ising pyrochlore magnets: Low temperature properties, 
        ice rules and beyond}

\author{R. Siddharthan$^{(a)}$, B. S. Shastry$^{(a)}$, 
         A.~P.~Ramirez$^{(b)}$, A.~Hayashi$^{(c)}$,
        R.~J.~Cava$^{(c)}$, S.~Rosenkranz$^{(d)}$} 
\address{$^{(a)}$Department of Physics, Indian Institute of Science, 
         Bangalore 560012, India}
\address{$^{(b)}$Bell Laboratories, Lucent Technologies, 600 Mountain Ave.,
          Murray Hill, NJ 07974}
\address{$^{(c)}$Chemistry Department, Princeton University, Princeton,
           NJ 08540}
\address{$^{(d)}$Materials Science Division, Argonne 
            National Laboratory, Argonne, IL 60436}
\date{March 15, 1999}
\maketitle

\begin{abstract}
Pyrochlore magnets are candidates for spin-ice behavior. We present
theoretical simulations of relevance for the pyrochlore family $R_2$Ti$_2$O$_7$
($R=$ rare earth) supported by magnetothermal measurements on selected systems.
By considering long ranged dipole-dipole as well as short-ranged superexchange
interactions we get three distinct behaviours: (i) an ordered doubly
degenerate state, (ii) a highly disordered state with a broad transition
to paramagnetism, (iii) a partially ordered state with a sharp transition to
paramagnetism.  Thus these competing interactions can induce behaviour very
different from conventional ``spin ice''. Closely corresponding behaviour
is seen in the real compounds---in particular Ho$_2$Ti$_2$O$_7$ corresponds
to case (iii) which has not been discussed before, rather than (ii) as
suggested earlier.
\end{abstract}
\pacs{PACS numbers: 05.50.+q, 75.10.Hk, 75.25.+z, 75.40.Mg, 75.50.Lk}

\bibliographystyle{unsrt} 


The pyrochlore rare earth titanates have attracted great
attention recently because their unusual structure (the
``pyrochlore lattice'') of corner-sharing tetrahedra can lead to
geometric frustration and interesting low-temperature properties
\cite{ramirez94}. Our interest in these particular titanates was
sparked by the observation (confirmed by our crystal field
calculations) that some of them are nearly ideal Ising systems
\cite{harris98prl,harris98jpcm,harris97}. Some intriguing
experimental data presented below can only be explained by
assuming a competition between classical dipole-dipole
interactions and quantum superexchange.  Depending on their
relative magnitudes, the ground states of the Ising-like systems
can be ``ice-like'', ordered, or partially ordered.  ``Ice
models'' get their name because real (water) ice
\cite{bernalpauling} has a large ground state degeneracy arising
from local rules for the ordering of protons in water ice.
Several related models have been studied since, but as far as we
know this is the first time that two competing interactions have
been included in such a model, with the physics changing
significantly depending on their relative strengths.

\begin{figure}[th]
\centerline{\epsfig{file=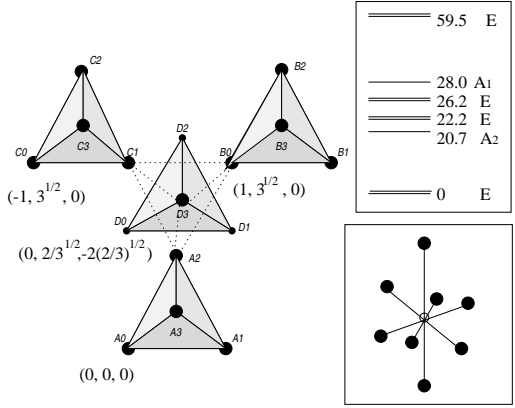, width=7.0cm, clip=}}
\caption{
\label{pyropic}
The basis of atoms $A0=(0,0,0)$, $A1$ $=$ $(r, 0, 0)$, $A2$ $=$ 
$r(1/2,\sqrt3/2,0)$, $A3$ $=$ $r(1/2,1/[2\sqrt 3],\sqrt{2/3})$; 
translated by the lattice
vectors 
${\bf a}_1$ $=$ $(r, \sqrt 3 r, 0 )$, 
${\bf a}_2$ $=$ $(-r, \sqrt 3 r, 0 )$, 
${\bf a}_3$ $=$ $(0, 2r/\sqrt 3, -2r \sqrt{2/3})$
to form tetrahedra $B$, $C$ and $D$;
repeated translation forms the whole lattice.
In our systems, $r=3.53$\AA. 
We can also choose a basis of ``downward''
tetrahedra (dotted lines; atoms $A2$, $B0$, $C1$, $D3$). ({\em inset bottom})
A single rare earth ion (centre) surrounded by eight oxygen ions. The
top two are at the centres of the tetrahedra adjacent to the rare
earth ion, and the rest form a puckered hexagonal ring around this
axis. ({\em
inset top}) The first few calculated energy levels for Ho, in meV, 
with symmetry indicated.
Ground state transitions are observed at 22, 26, 59, 71 and 77 meV (last two not
shown).
}
\end{figure}

Pyrochlores of the form $A_2B_2$O$_7$ have been extensively studied,
where $A$ are rare earth ions and $B$ are transition metal ions, each
forming interpenetrating pyrochlore lattices.
Often these can be modelled by isotropic Heisenberg antiferromagnets because
the $B$ atom is magnetic ($B$ = Mn, Mo) 
\cite{greedan96,gaulin92,greedan91} with a small dipole moment, 
and the $A$ atom is nonmagnetic (eg $A$=Y), so 
the dominant interaction between the $B$ atoms is superexchange.
The lattice is a three dimensional version of the kagom{\'e} lattice,
with a parallelepiped as the unit cell and a tetrahedron as the
basis (figure \ref{pyropic}). Typically the magnetic ions sit at the
corners of these tetrahedra. The tetrahedra form a face centred
cubic lattice, so the structure can be viewed as four interpenetrating
fcc lattices and the unit cell is often pictured as a cube, but we
used the smaller parallelepiped unit cell in our simulations.
The lattice can exhibit frustration; in the isotropic Heisenberg
case, this happens for the antiferromagnet \cite{gaulin92,moess98,reimers92}, 
but in an Ising limit it can happen even with ferromagnetic interactions
\cite{harris98prl,harris98jpcm,harris97}. 
 
In our systems, the Ti$^{4+}$, like the O$^{2-}$ ions, are nonmagnetic
and play no role apart from holding the
lattice together. However, typically the rare earth ion carries a
large magnetic moment (from its unfilled $f$-electron shells), so
that the dipolar interaction is as significant as the
superexchange.  
Another important aspect is the single ion anisotropy
imposed by the crystal field (CF) interaction of D$_{3d}$ symmetry at the
rare earth site, since a strong easy-axis
anisotropy results in the Ising limit, even for isotropic
exchange interactions. Previous investigations of the
low-temperature properties of these systems assumed
a strong single-ion anisotropy along the $\langle 111 \rangle$ direction,
i.e. along the line pointing
from the center of the tetrahedron to the corner where the rare
earth is located \cite{harris98prl,blote}
However, there is so far no direct experimental evidence to
support this assumption. We have therefore investigated
in detail the CF interaction
in the Ho-compound using inelastic neutron scattering
\cite{rosen}.

From the energies and intensities of the observed CF transitions,
we could unambiguously
determine the CF parameters and energy levels of 
Ho$_2$Ti$_2$O$_7$ (top inset to fig.~\ref{pyropic}). Because the crystal
structure varies very little on replacing one rare-earth ion by another,
these CF parameters will give good estimates of the splitting and
single ion anisotropy in the other compounds as well.
So we find a strong easy-axis anisotropy along the line joining
the tetrahedra centres for Ho and Dy, but not for Yb, Er or Tb.

Though it has been suggested earlier \cite{harris98prl,blote} that
Yb$_2$Ti$_2$O$_7$ is also Ising like, we find that there is in
fact an easy plane here, rather than an easy axis: $J=$ 7/2,
$J_z$ = $\pm 1/2$ for the ground states, so the spin points mainly
in the $x$--$y$ plane. The same seems to be true for Er, while
Tb may be Ising like but only at very low temperatures ( $<$ 0.1 K).

The nearest-neighbour Ising model on this lattice (considered in
\cite{harris98prl,harris98jpcm,harris97}) can show at most two kinds
of behaviour depending on the sign of the interaction.  If the
interaction is ``antiferromagnetic'', the ground state is doubly
degenerate and each tetrahedron has alternately all spins pointing out
or all spins pointing in.  If the interaction is ``ferromagnetic'',
the ground state of a tetrahedron is given by an ``ice rule'' where
two spins point out and two into the tetrahedron, and is sixfold
degenerate.  Any state with all tetrahedra satisfying this is a ground
state. It is highly degenerate with a finite entropy per spin, which
our simulations suggest is around 0.22 $k_B$ in agreement with
Pauling's prediction \cite{entropy}.  In both cases, the specific heat
vanishes at small as well as large temperatures, with a peak in the
middle. Simulations show that in the ferromagnetic case (ice rule) the
peak is broad, and occurs at the temperature scale of the interaction,
suggesting a typical broad crossover from a glassy low-temperature
phase with macroscopic entropy to a paramagnetic phase.  In the
antiferromagnetic case, the peak is very sharp and is at a temperature
around 4 times the interaction energy, suggesting a phase transition
from an ordered ground state to the paramagnetic phase.  The energy
scale of the peak here may be higher because the energy cost of a
single spin flip from the ground state is 12 times the interaction
energy of a pair of dipoles, as opposed to four times this energy in
the ferromagnetic case.

Experiments were done on polycrystalline samples of these compounds
which were synthesized  from stoichiometric mixtures of
the lanthanide oxides (99.99\%) and TiO$_2$ (99.995\%) heated at
1200$^\circ$ C in
air for 1 week with intermediate grindings. All materials were found to be
phase pure by conventional powder X-ray diffraction.
The specific heat was determined using a standard semiadiabatic
technique, and the susceptibility measured with a commercial magnetometer.
All susceptibility data were taken at 0.1 Tesla.

While simulations for the specific heat of 
the nearest neighbour model suggest a broad
crossover (antiferromagnetic interaction) or a sharp narrow peak
(ferromagnetic), for Ho something entirely different occurs: at around 0.6 K
a transition seems to occur, below which 
the spins seem to decouple thermally from the system and freeze
out into a low temperature metastable glassy phase. Moreover, the data
for Ho suggests a peak at substantially smaller energies
than the dipolar interaction (2.3~K).
To explain this we need to go beyond
the nearest neighbour model, by (a)~considering the long-ranged
dipole-dipole interaction between the spins, (b)~including an
antiferromagnetic superexchange to reduce the dipolar coupling.

It turns out that the Dy compound is very well described
by a purely dipole-dipole interaction but with a reduced effective
dipole moment (around 75\% of the full value). 
This could be explained by a superexchange which
falls off for the nearest few neighbours in roughly the same way as
the dipolar interaction. This compound is very interesting in its
own right, being a good realization of the ``ice models" which have
interested physicists for a long time, and we have discussed it
extensively elsewhere \cite{ramirez99}. 

The fact that Ho has significantly different behaviour from Dy means that
the superexchange behaves differently. It is not possible to account
for this cleanly, so we merely assume that the superexchange
is nearest-neighbour only: this still gives us excellent agreement with
the observations and highlights why these compounds are
different from ``spin ice''.
We calculate the dipole-dipole interaction,
assume a nearest-neighbour superexchange which we estimate from
the experimental data, do a simulation for the specific heat and
susceptibility with these values, and compare with experiment.
Our simulations are on systems with 2048 spins ($8\times 8\times
8$ tetrahedra each with 4 sites) and around 10000 Monte Carlo steps per spin.
We use a long-ranged dipole-dipole interaction (up to 5 nearest
neighbour distances, but the results don't change significantly beyond the
third neighbour). The convergence is good despite the long range of the
interaction, probably because there is no global Ising axis and no net
magnetization, so beyond the third neighbour the large numbers of spins
in different directions tend to cancel one another. 

We obtain the superexchange for Ho from the
experimental high temperature zero field susceptibility. The high
temperature expansion of the susceptibility is readily obtained
from elementary statistical mechanics. First we fix the notation:
We use scalar Ising spins, $S_i = \pm 1$, with $S_i=+1$ if it
points out of an ``upward'' tetrahedron (or, equivalently, into a
``downward'' tetrahedron'') and $S_i=-1$ otherwise. 
We write the first two terms in the expansion as
$\chi(T) = \frac{C_1}{T} \left( 1 + \frac{C_2}{T} \right)$
and try to evaluate these coefficients using 
$M = \frac{1}{N} g_s \mu_B \left< \sum_i S_i \cos \theta_i
\right>$
where $g_s$ is the Lande factor, $S_i$ is the effective spin of 
rare earth atom $i$ ($= \pm |J_z|$ for that atom), $\mu_B$ is the
Bohr magneton, and $\theta$ is the angle made by the
direction of the spin with the (arbitrarily chosen) direction of
the external magnetic field. Our results turn out to be
independent of the direction, at least to this order. 
The angle brackets denote the
thermodynamic average. From the fluctuation-dissipation
theorem,
$\chi(T) = \frac{1}{N} \beta (g_s \mu_B)^2 \sum_{i,j}
\Gamma_{ij}$
where 
$\Gamma_{ij} = \left< S_i \cos \theta_i~ S_j \cos \theta_j
\right>_{{\cal H}=0}$.  Using standard
methods (expanding to order $\beta$), we
arrive at
\begin{eqnarray}
\chi(T) & = & \frac{N(g_s \mu_B)^2}{k_B T} \frac{S^2}{3} 
        \left[ 1-  \frac{6 S^2}{k_B T} \frac{1}{4}
         \sum_{i \atop {\rm {over~ 1}\atop tetrahedron}} \sum_j J_{ij} \cos \theta_i
                \cos \theta_j \right]  \nonumber
\end{eqnarray}
The sum over $j$ is over all sites in the lattice excluding $i$. 
If the nearest-neighbour dipolar interaction is $J_D$, the superexchange
is $J_S$, and we include long-ranged dipolar interaction but only nearest
neighbour superexchange, we get
\[
\chi(T) = \frac{N (g_s \mu_B)^2}{k_B T} \frac{S^2}{3} \left[
               1 + \frac{6 S^2}{k_B T} \frac{1}{4} \left( 2.18 J_D + 2.67 J_S
         \right) \right]
\]
and from here we can extract the coefficients $C_1$ and $C_2$.

\begin{figure}[th]
\centerline{\epsfig{file=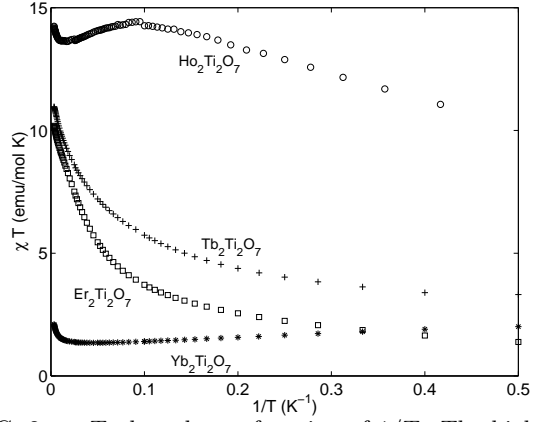, width=7.0cm, clip=}}
\caption{ \label{htsusc}
$\chi T$ plotted as a function of $1/T$.
The high temperature expansion in the text is the markedly
linear low temperature
region here (2--10 K, which is high compared to $C_2$).
Note that the Yb compound has
the opposite slope here from Ho, suggesting that
superexchange dominates here.
}
\end{figure}

This is valid for an ideal Ising model at sufficiently
high temperatures. When we plot the experimental $\chi T$
against $1/T$ (Fig.\ \ref{htsusc}), 
we find a marked linear region at low temperatures
(2--10 K).
This is the region we want: if we pull out $C_2$ from
this region, we find it is much less than 1 K, so things are
consistent. At higher temperatures, where the Ising approximation
should fail, the graph is no longer linear.
We equate this value of
$C_2$ to $(6S^2/4)(2.18J_D+2.67J_S)$ with $J_D$ known, pull out $J_S$,
and do the simulation. In fact, since the slope is so
small, the error is not too great if we simply put $C_2=0$. But using
the measured slope of $C_2$ (and using the calculated value of $C_1$,
for consistency, rather than the fitted value) we get $J_D = 2.35$ K
(calculated) and $J_S = -1.92$ K (measured), both for the Ho and the
Dy compounds. (Note that when we use scalar Ising spins rather than
fixed vector spins, the superexchange is {\em negative}
and the dipolar $J$ is {\em positive}---and the former favours ordering,
the latter frustration, as is usually the case in Ising systems.)

\begin{figure}
\centerline{ \epsfig{file=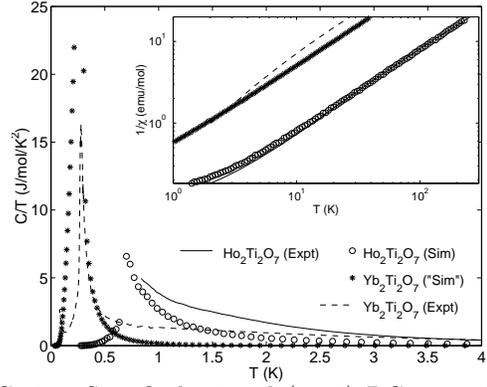,width=7.0cm, clip=}}
\caption{ \label{hoyb}
Specific heat and (inset) DC susceptibility
for Ho$_2$Ti$_2$O$_7$ and Yb$_2$Ti$_2$O$_7$. The Yb ``simulation'' here is
for an Ising model, which is probably inappropriate but gives fair agreement. 
}
\end{figure}

We now simulate with these values of $J_D$ and $J_S$.
In the case of Ho$_2$Ti$_2$O$_7$ (Fig.\ \ref{hoyb}),
the simulated susceptibility agrees well with the experimental
data at all temperatures, while the specific heat has a sharp peak
at very nearly the point where the experimental Ho system
falls out of thermal equilibrium. Moreover, there is a large
energy difference at this point, suggesting a first-order phase
transition.

The Yb compound has earlier been believed to be an Ising model,
and we initially tried modelling it in this way, with a nearest-neighbour
antiferromagnetic superexchange. The experimental data show a
sharp peak, which is as we expect for an antiferromagnetic
Ising model, and matching the position of the peak in the
simulation in the observed position leads to fair 
agreement with experiment (fig.\ \ref{hoyb}). 
This should be regarded as fortuitous.
The neutron data suggest that the Yb and Er compounds are
easy-plane (``XY models''), not Ising. 
Earlier work by Bramwell {\em et al.\ }\cite{bramwell94}
suggests that the XY Heisenberg model on this lattice shows
a first order phase transition from an ordered ground state;
we believe that, as with Ho$_2$Ti$_2$O$_7$, it may be necessary
to include a dipole-dipole interaction, and preliminary simulation
of a pure dipole model correctly predicts the position and
approximate shape of the peak.  More work on this is in progress.
Our specific heat measurements on Er and Yb agree with previous
data \cite{blote}.

\begin{figure}[th]
\centerline{\epsfig{file=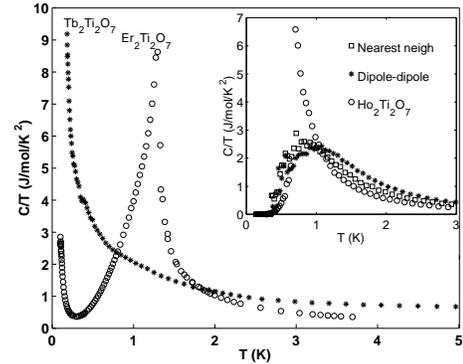,width=7cm,clip=}}
\caption{\label{tber}
The measured specific heat for Tb$_2$Ti$_2$O$_7$ and Er$_2$Ti$_2$O$_7$,
and (inset)
a comparison of three simulations for the specific heat: the
nearest neighbour dipolar interaction, the long ranged
dipolar interaction, and the long-ranged dipolar interaction
modified by a nearest neighbour superexchange (which we
use for the Ho compound). The third
case is very different from the first two. 
}
\end{figure}

The remaining compound, Tb$_2$Ti$_2$O$_7$, is probably Ising-like
at very low temperatures. It has been suggested that it remains
paramagnetic down to 0.07 K \cite{gardner}. The gap to the excited CF states is
only a few kelvin. The data for this and Er are shown in fig.\ 
\ref{tber}, but no simulations were done for these.

The ground states of nearest-neighbour ferromagnetic or
antiferromagnetic Ising pyrochlores are well known;
we now consider the more complicated case of Ho$_2$Ti$_2$O$_7$.
In the nearest neighbour ferromagnetic
model any state in which all tetrahedra satisfy the 
``ice rule'' will be a ground state. With long-ranged
interactions (Ho$_2$Ti$_2$O$_7$), the ice rule 
remains but there are further restrictions
on the allowed ground states. One way to deal with these restrictions
is to consider the system as a set of interacting ``upward'' tetrahedra,
assign ice rules to the configurations of each, and determine what
configuration of neighbouring tetrahedra will lower the energy. There
should be an ice rule for the ``downward'' tetrahedra, but also
other constraints.  This will map on to a new
interacting lattice system, with a six valued variable at each lattice
site representing the configuration of the upward tetrahedron represented
by that lattice site.

This approach for Ho, and the simulation,
suggest a partial ordering in the ground state. That is, the
upward tetrahedra could have one of two allowed configurations;
the configuration varies randomly along one lattice direction, 
but alternates perfectly along the other two. 
So the number of ground states is large
but not macroscopic (it is exponential in $L$, the system length,
rather than $L^3$), and the entropy per particle vanishes. 
Our calculation here ignored long-ranged superexchange, and thus
may not be valid for the experimental system, but the experimental
data for Ho do suggest a vanishing entropy for the ground state
(on integrating $C/T$).
In the simulation, the system remains in a disordered
``paramagnetic'' state till the transition temperature, but below
this temperature it freezes out rapidly to such a
partially ordered state. From then on further cooling leaves it
stuck in this state, with the other ground states inaccessible.
This seems to agree with the observation that the spins
freeze in Ho$_2$Ti$_2$O$_7$ below a temperature of around 0.6 K. 
Below this temperature inability to establish thermal equilibrium
leads to unreliable data, which has not been plotted here.

The freezing of spins is an interesting phenomenon in Ho$_2$Ti$_2$O$_7$.
We believe it is because at the transition temperature (0.8 K)
the single spin flip energy is around 4 K, so the Boltzmann
factor for this is very small (around 0.006). A single spin flip
from the ordered ground state for Yb has a much larger Boltzmann factor of
0.05 (assuming only near-neighbour interaction), so the fall in
specific heat is not so sharp.
This ``spin freezing'' in Ho$_2$Ti$_2$O$_7$ has been 
commented on by Harris {\em et al.}.
It seems the next-neighbour interactions must be fairly strong for this.
In the absence of next-neighbour interactions there is a very large number of
ground states. As we turn on and increase the next-neighbour
interaction, the new constraints substantially reduce the ground
state entropy. For a pure dipole-dipole interaction the new 
ground state entropy
is reduced but still finite, but with large superexchange, 
as in Ho$_2$Ti$_2$O$_7$, it actually
vanishes: there are few true ground states, and these are
separated by large energy barriers.

In summary, we perform simulations based on a theoretical
calculation of dipole-dipole interactions and an estimated
superexchange obtained from the experimental data. 
The relative strengths of these interactions have a drastic
effect on the ground state properties when compared to a
nearest-neighbour Ising model,
and we observe three different kinds of ground states: 
highly disordered and ice-like \cite{ramirez99}, partially ordered, or
fully ordered, with broad crossovers or sharp phase
transitions to high temperature phases. 
We use only one adjustable parameter,
fitted from the experimental data, as input for the simulations 
that agree well with experiment.
Thus these systems look like excellent testing grounds
to study the behaviour of disordered spin systems, glassy
dynamics, and phase transitions, with the opportunity to 
tune the interactions to some extent, and should
richly repay future study.

We thank C.~Dasgupta for helpful
discussions. SR's work was supported by US DOE BES-DMS W-31-109-ENG-38.

\twocolumn

\begin{thebibliography}{88}
\bibitem{entropy} 
Pauling's argument, originally for water ice, applies here as follows: 
We can give ice rule
configurations to the ``upward'' tetrahedra in $6^{N/4}$ ways. 
But not all of these are valid ground states:
the $N/4$ ``downward'' tetrahedra,
which are formed by the corners of these ``upward'' tetrahedra,
should also satisfy the ice rule. The probability that a 
random tetrahedron satisfies the ice rule is $6/16$; so
we deflate the above number by this factor for
each downward tetrahedron, and obtain
$6^{N/4} (6/16)^{N/4}$ states, and an entropy per spin 
of 0.20---within 10\% of the simulation.
\end{thebibliography}
\end{document}